\documentclass[a4paper,11pt]{article}
\usepackage{jinstpub} 
\usepackage{lineno}
\usepackage{graphicx}
\usepackage{subcaption}
\usepackage[T1]{fontenc}
\usepackage[utf8]{inputenc}
\usepackage[normalem]{ulem}
\usepackage{booktabs} 

\makeatletter
\renewcommand{\proceeding}[1]{%
  \gdef\@proceeding{#1\vspace{1.5cm}}%
}
\makeatother

\proceeding{TWEPP 2025 Topical Workshop on Electronics for Particle Physics October 4, 2025 to October 10, 2025 \\ Rethymno, Crete, Greece}

\title{A Latency-Constrained, Gated Recurrent Unit (GRU) Implementation in the Versal AIE}

\author{M. Sapkas$^1$,\note{Corresponding author.}}
\author{A. Triossi}
\author{and M. Zanetti}
\affiliation{Università e INFN, Padova,\\ Via Marzolo, 8 - 35131, Padova, Italy}

\emailAdd{michail.sapkas@phd.unipd.it}

\abstract{This work explores the use of the AMD Xilinx Versal Adaptable Intelligent Engine (AIE) to accelerate Gated Recurrent Unit (GRU) inference for latency constrained applications. We present a custom workload distribution framework across the AIE's vector processors and propose a hybrid AIE - Programmable Logic (PL) design to optimize computational efficiency. Our approach explores the parallelization over the rows of the matrices by utilizing as many of the AIE vectorized processors effectively computing all the elements of the resulting vector at the same time, an alternative to cascade stream pipelining.}


\notoc
\begin{document}
\maketitle
\flushbottom

\section{Introduction}
\label{sec:intro}

With the introduction of the \textit{Versal} family, AMD presents the \textit{Adaptive Compute Acceleration Platform} (ACAP) \cite{amd_versal_adaptive_soc}. One of the key innovations within this platform is the \textit{Adaptable Intelligent Engine} (AIE), a specialized accelerator array optimized for high throughput multiply accumulate (MAC) operations on floating point and fixed point vector data. We acknowledge prior research efforts that efficiently map generalized matrix matrix (GEMM) operations onto the AIE, such as \textit{MaxEVA}~\cite{taka2023maxevamaximizingefficiencymatrix} and \textit{CHARM}~\cite{zhuang2023charmcomposingheterogeneousaccelerators}. However, these works primarily target throughput optimization in synthetic matrix multiplication scenarios, whereas our focus is on real time inference under stringent latency constraints. Furthermore, Recurrent Neural Networks (RNNs) are notoriously difficult to parallelize efficiently on hardware, making their low latency implementation a challenging yet valuable objective. In this work, we present a proof of concept implementation of a latency constrained GRU on the Versal AIE. The design explores several unconventional aspects of the architecture, including the use of free running kernels \cite{amd_aie_free_running_ug1079}, row-wise matrix vector operations for efficient parallelization across hidden state dimensions, and a novel utilization of the AIE’s interface tiles \cite{amd_pl_interface_tile_capabilities_ug1079} as dynamic data aggregators. A direct comparison with GPU/CPU baselines is outside the scope of this
work, which focuses on architectural trade offs within Versal.
\section{Recurrent Neural Networks - Gated Recurrent Unit}
\label{sec:rnns_gru}
RNNs \cite{DBLP:journals/corr/abs-1912-05911} are a specialized class of neural architectures designed to process sequential or time dependent data and incorporate temporal dependencies by maintaining a hidden state that evolves over time. 
\begin{figure}[h!]
    \centering
    \begin{subfigure}[h!]{0.4\textwidth}
        \centering
        \includegraphics[width=\linewidth]{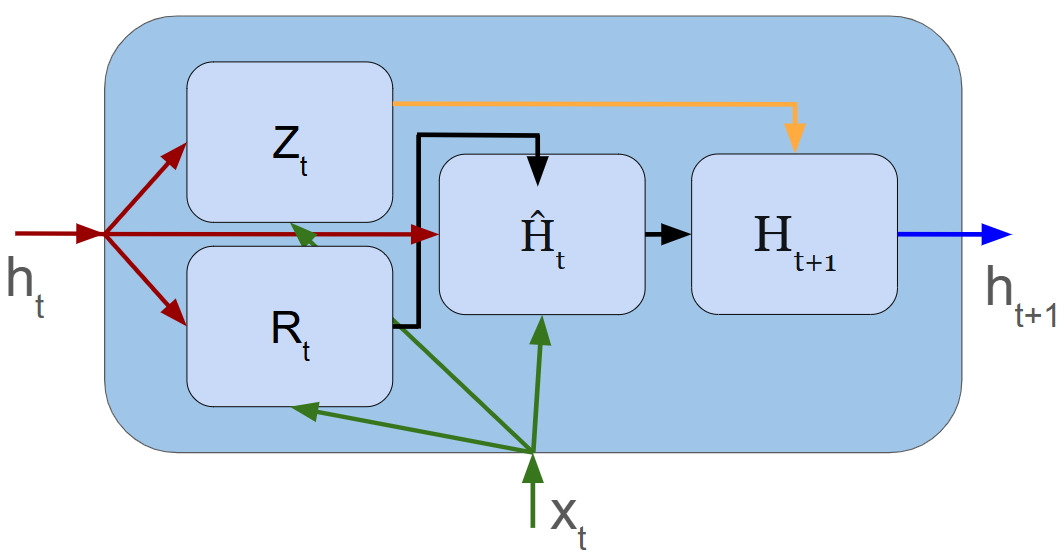}
        \caption{Data flow between gates inside the GRU "cell".}
        \label{fig:rnns_gru}
    \end{subfigure}
    \hfill
    \begin{subfigure}[h!]{0.42\textwidth}
        \centering
        \includegraphics[width=\textwidth]{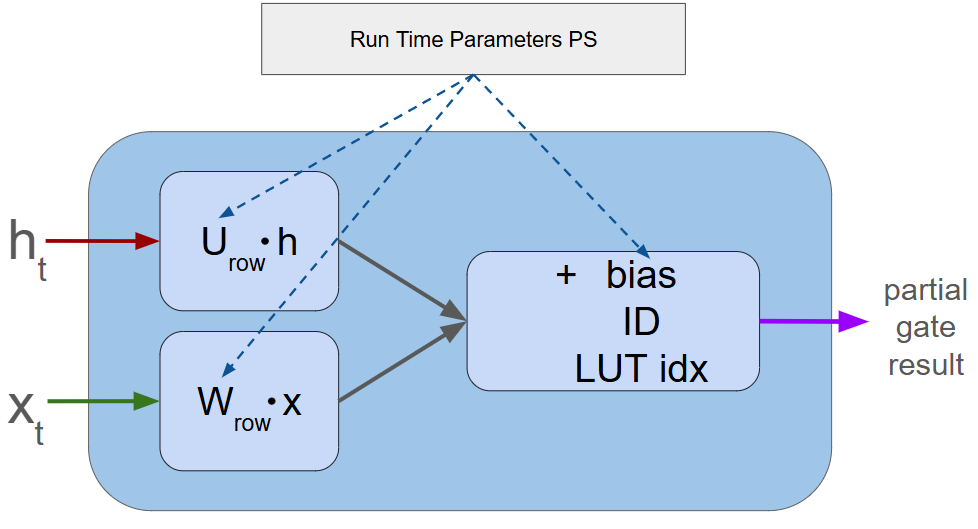}
        \caption{Compute one element of a gate.}
        \label{fig:temp_gru}
    \end{subfigure}
    \caption{\textbf{(a)} A forward pass of the GRU. \textbf{(b)} Each of the \(Z_t\), \(R_t\), \( \tilde{H}_t \) gates partial result needs three AI tiles to compute, each consuming a row of the matrix and the whole vector. The third tile adds the results together with the bias, assigns an ID, and transforms the floating point to a LUT index to be sent to the PL kernel (GRU HYBRID), or applies the LUT in the AIE and then sends it to a specialized aggregator/sorting AIE kernel via the packet merge construct (GRU AIE). }
    \label{fig:gru}
\end{figure}
The Gated Recurrent Unit (GRU) \cite{cho2014learning}, along with the Long Short-Term Memory (LSTM) \cite{hochreiter1997}, represents one of the most widely used RNN variants and use gating mechanisms as seen in the GRU Cell in Figure~\ref{fig:rnns_gru}. These gates are implemented via linear transformations followed by nonlinear activation functions. The Reset gate (\( R_t \)) determines how much of the previous hidden state should be forgotten, while the Update gate (\( Z_t \)) controls how much of the Candidate New Hidden State (\( \tilde{H}_t \)) should contribute to the current hidden state. Mathematically, the GRU can be expressed as:
\[
\begin{aligned}
Z_t &= \sigma(W_z x_t + U_z h_{t-1} + b_z) \\
R_t &= \sigma(W_r x_t + U_r h_{t-1} + b_r) \\
\tilde{H}_t &= \tanh(W_h x_t + U_h (r_t \odot h_{t-1}) + b_{\tilde{h}}) \\
H_t &= (1 - z_t) \odot h_{t-1} + z_t \odot \tilde{h}_t
\label{eq:gru}
\end{aligned}
\]
where \( \sigma(\cdot) \) denotes the sigmoid activation, \( \tanh(\cdot) \) the hyperbolic tangent, and \( \odot \) the element-wise (Hadamard) product. Because each timestep depends on the previous hidden state, GRUs remain inherently sequential in computation, limiting parallelization. Moreover, within a single timestep, dependencies among the gates impose additional sequential constraints. Typical latencies for a forward pass on GPUs are in the order of milliseconds \cite{lstm_lat_gpu}.

\section{The Versal Adaptable Intelligent Engine on the VCK190 Platform}
\label{sec:aie}

 The AIE is a two dimensional tiled array composed of 400 RISC-V vector processors, interconnected through a high bandwidth network on chip. Data exchange between AIE and PL occurs via 39 specialized interface tiles that employ the AXI4-Stream protocol. These interfaces convert 128 bit data streams from the PL domain to 32 bit streams suitable for the AIE domain. Given that the AIE operates at a maximum clock frequency of 1.25~GHz, achieving peak throughput requires the PL domain to be clocked at one quarter of this frequency (312.5~MHz) to maintain synchronization. These are also the frequencies that we used for our implementation. Each AIE tile features a vector processor coupled with 32~KB of local memory. The vector unit supports both integer and floating point arithmetic, using two distinct execution pipelines. It can perform up to two 128 bit load operations and one store operation per clock cycle. To exploit this vectorized architecture, developers use C++ intrinsics or, more conveniently, the high level APIs provided in recent AMD toolchains~\cite{AIE_API_2024}. Local memory within a tile typically serves as stack and heap storage, as well as a ping pong buffer for efficient double buffered data transfers. A particularly useful feature of the AIE architecture is the support for \emph{runtime parameters}, which allow values to be dynamically loaded into the memory banks of specific AIE tiles by the processing system (PS). AIE tiles can communicate with one another through four principal mechanisms, two local: Cascade Streams and Input/Output Buffers, and two non local: AXI4-Streams and Packet Streams. We will make use the two latter. One of the main advantages of the AIE over traditional programmable logic is its native support for floating point arithmetic, which eliminates the need for quantization when deploying AI models. In addition, AIE computation is deterministic and does not require timing closure, a common challenge in FPGA design.

\section{Model Implementation}
\label{sec:impl}
Our implementation aims to explore the capabilities of the AIE as a real time inference device \textbf{using the Single Precision Floating Point (32 bit) data type}. In this context, a forward pass of the model processes a \textbf{single batch}, and all AIE kernels execute within an \textbf{infinite} \texttt{while} \textbf{loop}. To implement the matrix vector multiplication, we leverage the vectorized Multiply Accumulate (MAC) operation available within each AIE tile. In this process, the input or hidden state vector is streamed into the vector registers, while the matrix parameters are loaded from the tile’s local memory. The efficiency of this computation depends primarily on two factors: the data streaming bandwidth and the MAC pipeline. There are two principal approaches for executing this operation:

\medskip \noindent \textbf{1. Column-wise Cascade:} This method represents the most hardware compliant, or “lawful,” implementation in that it fully exploits the AIE’s most efficient data transfer mechanism; the cascade stream. The matrix parameters are partitioned into column blocks, and each tile performs MAC operations between ablock of matrix columns and the corresponding elements of the input vector (which are broadcasted to match the vector width). The partial results of eachblock are then cascaded to the neighboring tile and added. The number of rows processed in each operation corresponds to the number of vector lanes of the chosen data type and each participating tile must process all the rows of the matrix which can create memory limitations.
\begin{figure}[h!]
    \centering
    \begin{subfigure}[b]{0.3\textwidth}
        \centering
        \includegraphics[width=\textwidth]{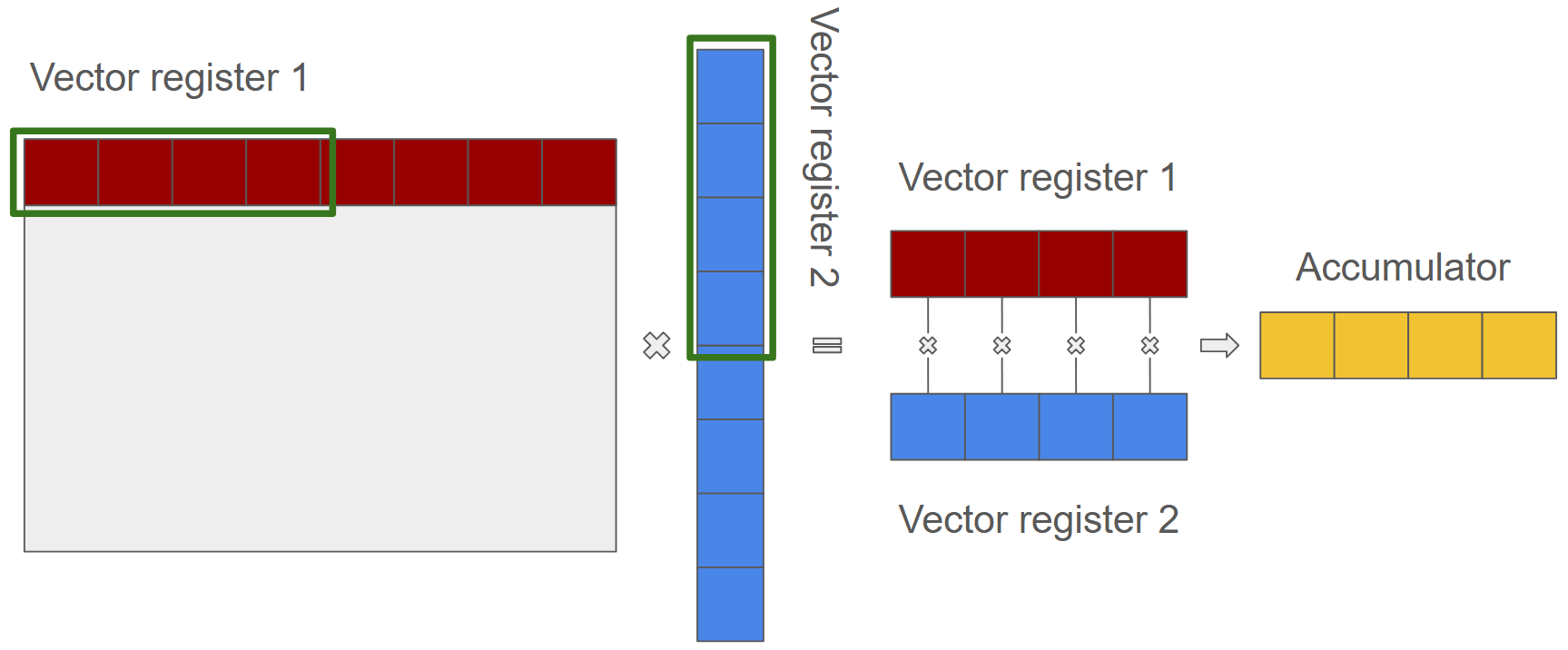}
        \caption{MAC the first part of the row with first part of the vector, store the result.}
        \label{fig:row1}
    \end{subfigure}
    \hfill
    \begin{subfigure}[b]{0.3\textwidth}
        \centering
        \includegraphics[width=\textwidth]{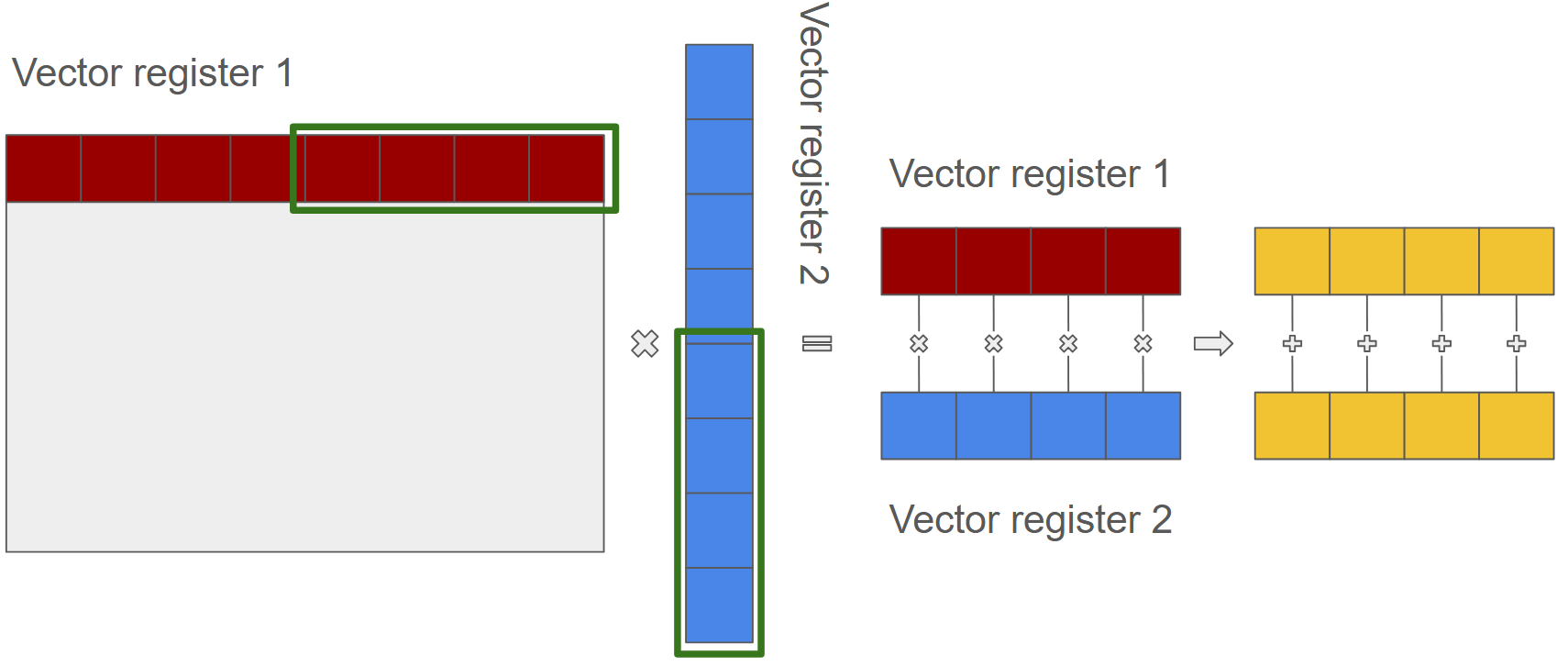}
        \caption{Continues with the rest of the row with rest of the vector, accumulate on previous result.}
        \label{fig:row2}
    \end{subfigure}
    \hfill
    \begin{subfigure}[b]{0.35\textwidth}
        \centering
        \includegraphics[width=\textwidth]{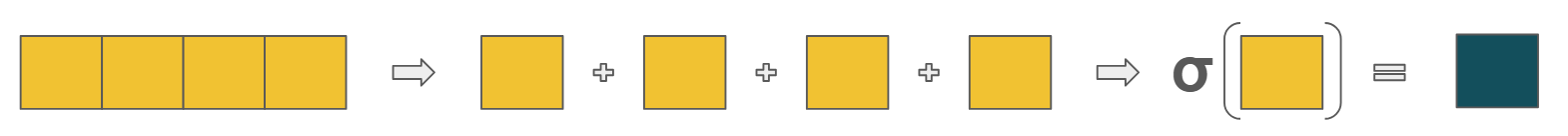}
        \caption{When the whole row and vector have been consumed, reduce the vector, Finally, apply activation function, the element that corresponds on the row of the matrix (here is the first) of the resulting vector is computed.}
        \label{fig:red}
    \end{subfigure}
    \caption{Row-wise implementation of the matrix vector multiplication within the AIE.}

    \label{fig:mult}
\end{figure}

\medskip
\noindent \textbf{2. Row-wise Streams:} This method addresses the limitations of the column-wise cascade by transposing the computation. Instead of performing the MAC operation along the columns, we now accumulate across the rows of the matrix. In this configuration, each tile MACs across an entire matrix row, progressing through the columns at a width determined by the vector lane widthas shown in Figure~\ref{fig:mult}. This approach accelerates traversal through the matrix but requires thewhole vector to be streamed to all participating tiles, rather than partitioned into smaller segments. In hindsight, this introduces certain inefficiencies in data movement since we cannot stream fast enough to exploit the best possible pipeline frequency of a tile. Thankfully, this limitation applies only for the first rows computed in a tile. All next rows can benefit by the vector being kept in memory and now we are only limited by how fast we can load from memory. To distribute the full vector efficiently, we exploit the streaming interface’s capability to broadcast the same data stream from a single interface tile to multiple destinations. With this broadcast mechanism incurs a latency penalty. A data flow of three AIE tiles, as shown at Figure~\ref{fig:temp_gru} was used as template for the matrix vector multiplications of the Update and Reset gates. An AI Tile performs the \( \mathbf{W_{row}} \cdot x \) while another computes \( \mathbf{U_{row}} \cdot h \). A third tile accepts the dot product results, adds the bias, incorporates an ID to the packet and either sends it to the PL or applies a LUT for the activation function. Specialized \(\tilde{H}_t\) AIE kernels are needed to interface with the R gate. This strategy allows for extensive utilization of AIE resources, enabling multiple matrix rows to be processed concurrently and using streams to move only one 32b numbers. Furthermore, it introduces the possibility of a “row reuse” factor, wherein each tile computes several consecutive rows. Using the template above the following formula calculates the AI tiles used for a single GRU layer: \( 3 \times HIDDEN \times 3 + 1  \). It corresponds to three AI tiles, for every hidden state dimension, for three gates,  plus one AI Tile for the New Hidden State Gate (\(H_t\)). In case of GRU AIE architecture another three kernels are added as aggregators.

\medskip
\noindent \textbf{Aggregation of Partial Results}

The distributed partial results are then aggregated to reconstruct the output vector. One approach is to perform this reduction directly within the AIE using packet streams combined through the \texttt{merge} construct. However, the arrival order of the packets is difficult to predict before implementation. To manage this, we assign unique identifiers to each packet. These IDs are then used to sort the packets, reconstruct the output vector, and broadcast it to the subsequent processing stage (e.g., the next GRU gate). While functionally correct, this approach breaks pipeline continuity: downstream AIE kernels must wait for the entire vector to be reassembled before proceeding, thereby introducing a latency bottleneck. This leads to a natural question: can aggregation be achieved faster? 
We introduce a “hybrid” solution in which the interface tiles are used as a data path to a PL aggregation kernel. To the best of our knowledge, this is the first application of multiple interface tiles in such a role for a machine-learning workload. In this configuration, the partial results are streamed directly from the AIE into the PL fabric. From the AIE side, each interface tile connects to a three way packet-merge construct, allowing partial results from all three GRU gates to share the same interface tile.  The corresponding PL kernel, implemented in High Level Synthesis (HLS)~\cite{UG1399_Vitis_HLS_2024}, operates as a finite state machine that executes the following sequence: 
1) Perform blocking reads from all interface tiles; 
2) Decode packet identifiers, extract LUT indices for the activation function; 
3) Apply the appropriate activation function; 
4) Write the output to the appropriate interface tile, padding with zeros when necessary to meet hardware alignment requirements.The timing and resources results for the PL kernel can be found in Table~\ref{tab:1}. The column "AIE AGGR TILE LAT" measures only the latency of the specialized aggregator AI tile, to serve as a comparison with the PL kernel.

\begin{table}[h]
\centering
\small
\resizebox{\textwidth}{!}{%
\begin{tabular}{c c c|cccc|c}
\cline{4-7}
 &  &  & \multicolumn{4}{|c|}{PL kernel} &  \\
\cline{4-7}
\shortstack{INTERFACE TILES \\ HIDDEN DIM} &
\shortstack{AIE GRU \\ TILES USED} &
\shortstack{AIE HYBRID \\ TILES USED} &
II &
DEPTH &
FF &
LUTS &
\shortstack{AIE AGGR\\TILE LAT} \\
\hline
20 & 184 & 181 & 6 & 7 cc = 22.4 ns & 5124 & 9250  & 229.6 ns \\
24 & 220 & 217 & 6 & 7 cc = 22.4 ns & 6066 & 11834 & 273.6 ns \\
28 & 256 & 253 & 8 & 9 cc = 28.8 ns & 7057 & 15663 & 319.2 ns \\
32 & 292 & 289 & 8 & 9 cc = 28.8 ns & 8003 & 19198 & 363.2 ns \\
\hline
\end{tabular}%
}
\caption{Resource utilization and timing results for the PL kernel and the AIE}
\label{tab:1}
\end{table}

\section{Latency Measurements and Conclusions}
We focus exclusively on the row-wise implementation. Latency measurements are performed using Single Precision (32b) Floating Point (FP) 8-lane Vectors \cite{AMD_UG1079_VectorDataTypes} on a single-layer GRU with hidden size up to 32 units. We are processing a single input vector per iteration (batch size  \(= 1\)), using synthetic inputs and weights to characterize hardware latency independently of
model accuracy. In order to scale the hidden and input sizes, it is not possible to use only one dataset, but we numerically tested the \(H=20\) and \(X=5\) with a GRU trained in a jet tagging dataset \cite{khoda2022ultralowlatencyrecurrentneural}. In this application, the GRU exhibits a computational dependency on its previous hidden state and thus, the overall latency of the model and the iteration interval should be identical, since each iteration must await the completion of the preceding one. The AI tiles responsible for performing the \( \mathbf{W} \cdot x \) matrix vector multiplication are decoupled from the rest of the computational pipeline, allowing them to prefetch and process new input data as long as they do not experience backpressure from downstream tiles.
\begin{figure}[h!]
  \centering
  \includegraphics[width=0.92\textwidth]{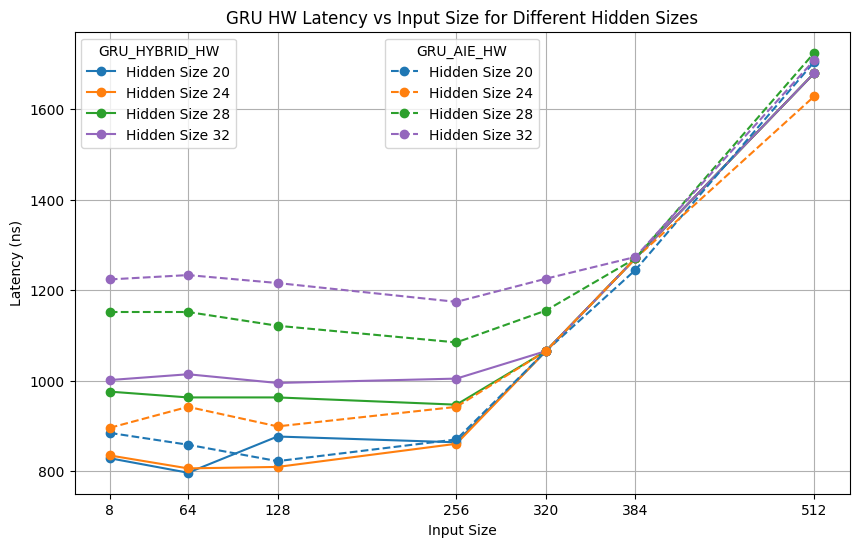}
  \caption{Latency results for a GRU using the PL kernel called "Hybrid" and the GRU implementation completely inside the AIE using specialized aggregation/sorting AI Tiles called "AIE".}
  \label{fig:results}
\end{figure}
Latency measurements are performed directly in hardware. We monitor two consecutive \texttt{TVALID} signal activations on the selected output stream interface, measuring the elapsed time between the assertion of the signal for the first valid datum and its subsequent activation for the next iteration. This interval corresponds to the true latency of a complete forward pass through the model. The key features of Figure~\ref{fig:results} are: 1) The better latency of the Hybrid model scaling up with the hidden state dimensionwhich is \(163~ns\) and \(197~ns\) on average for hidden sizes 28 and 32. 2) The latency plateau with respect to the input size showcases the decoupling of the \( \mathbf{W} \cdot x \) computing kernels. The dot product is in the shadow of the data transfers but after input dimension 256 it adds nanoseconds linearly as it starts dominating in latency. Finally, we suffer limitations in the number of available packet merge connectionsfor the GRU AIE architecture (32). The GRU HYBRID architecture is also constrained by the number of interface tiles (39). In case we need to compute larger hidden states, the same architecture can be used by allowing the tiles to compute more matrix rows and reuse the same data transfer and aggregation paths, computing 32 outputs at a time. Therefore we expect that the PL kernel will accumulate latency gains over the AIE packet merge in the regime \(H = 32\) and \(X = [8, 256] \) .

\bibliographystyle{JHEP}
\bibliography{biblio}

\end{document}